\documentclass[12pt]{article}
\usepackage{amsmath,amssymb}
\usepackage{epsfig} 

\newcommand{\newc}{\newcommand}
\newc{\gev}{\hbox{\rm\,GeV}}
\newc{\tev}{\hbox{\rm\,TeV}}

\newc{\gsim}{\lower.7ex\hbox{$\;\stackrel{\textstyle>}{\sim}\;$}}
\newc{\lsim}{\lower.7ex\hbox{$\;\stackrel{\textstyle<}{\sim}\;$}}

\def\simlt{\stackrel{<}{{}_\sim}}

\def\eq#1{eq.~(\ref{#1})}

\def\slash#1{\not  \! \! {#1}}

\def\vev#1{\langle {#1} \rangle}
\def\identity{1 \!{\rm l}}
\def\beq{\begin{equation}}
\def\eeq{\end{equation}}
\def\bea{\begin{eqnarray}}
\def\eea{\end{eqnarray}}

\begin{document}

\baselineskip=18pt

\setcounter{footnote}{0}
\setcounter{figure}{0}
\setcounter{table}{0}

\begin{titlepage}
\begin{flushright}
CERN-PH-TH/2008--069\\
\end{flushright}
\vspace{.3in}

\vspace{.5cm}

\begin{center}

{\Large\sc{\bf Higgs-dependent Yukawa couplings}}

\vspace*{9mm}
\renewcommand{\thefootnote}{\arabic{footnote}}

\mbox{ \bf Gian F.~Giudice and Oleg Lebedev}

\vspace*{0.9cm}

{\it CERN, Theory Division,  CH--1211 Geneva 23, Switzerland}

\end{center}

\vspace{1cm}

\begin{abstract}
\medskip
\noindent
We consider the possibility that the Yukawa couplings depend on the Higgs field, with the motivation of generating the fermion mass hierarchy through appropriate powers of the Higgs vacuum expectation value. This leads to drastic modifications of  the Higgs branching ratios, new Higgs contributions to various flavor-violating processes, and  observable rates for the top quark decay $t\to hc$. The underlying flavor dynamics must necessarily appear at the TeV scale and  is within the reach of the LHC.

\end{abstract}

\bigskip
\bigskip

\end{titlepage}


\section{Introduction} 
\label{sec_intr}

The theoretical understanding  of fermion masses and mixing angles is one of the most important unresolved problems in particle physics. In this paper we will try to address this problem by introducing an effective theory, valid below a new mass scale $M$, in which the Yukawa couplings depend upon the Higgs field $H$. The theory does not contain any small coupling constants and  the hierarchical pattern of the fermion masses is understood in terms of powers of $\vev{H}/M$. The new mass scale $M$ turns out to be necessarily around  the TeV, suggesting a possible link between flavor dynamics and the physics associated with electroweak symmetry breaking.  A similar proposal was made in 
ref.~\cite{Babu:1999me}.

At first sight it may seem that introducing flavor dynamics at the TeV scale will have disastrous consequences for flavor-changing neutral current (FCNC) processes. Indeed, in the effective theory, new contributions to FCNC are already induced at tree level by Higgs-boson exchange. Nevertheless, as we will show in sect.~\ref{sec_fla}, the effect is sufficiently small, since the flavor-violating Higgs couplings are related to the pattern of fermion masses.
An explicit example of how Higgs-dependent Yukawa couplings can be generated at the scale $M$ is presented in sect.~\ref{sec_mod}.

Since the scale of flavor dynamics $M$ is linked to the electroweak scale, our approach leads to many experimentally-observable distinctive features. New flavor-violating effects are expected just beyond the present experimental bounds. The new states at the scale $M$ are well within the reach of the LHC. Moreover, as we  describe in sect.~\ref{sec_hig}, the Higgs branching ratios are drastically modified, because the Higgs couplings to fermions are larger than those in the Standard Model (SM) by a factor of a few. Finally, the Higgs boson can decay in flavor-violating channels or, depending on its mass, be produced in the top-quark decay $t\to hc$.

\section{The effective theory of Higgs-dependent Yukawa couplings}
\label{sec_eff}

Our basic assumption is that, in an effective theory valid below the new-physics scale $M$, quark and lepton Yukawa couplings $Y^{u,d,\ell}$ are functions of the Higgs field $H$. 
Therefore, the Yukawa couplings can be expanded as
\beq
Y_{ij}(H)= \sum_{n=0}^\infty c_{ij}^{(n)} \left( \frac{H^\dagger H}{M^2}\right)^n,
\label{yukexp}
\eeq
where $i,j$ are generation indices. In what follows we focus upon the particularly interesting possibility that the hierarchical pattern of fermion masses is explained by appropriate powers of $\vev{H}/M$. Therefore, we assume that the coefficients $c_{ij}^{(n)}$ vanish up to a (generation-dependent) order $n_{ij}$. Keeping only the  leading order terms 
 in \eq{yukexp}, we can express the Yukawa effective Lagrangian as
\beq
 -{ \cal L}_Y=    Y^{u}_{ij}(H) ~   \bar q_{Li} u_{Rj}  H^c
+      Y^{d}_{ij}(H) ~     \bar q_{Li} d_{Rj}  H+{\rm h.c.}~, 
~~~~~~~~~~~Y^{u,d}_{ij}(H)=c_{ij}^{u,d}  \left( {H^\dagger H \over M^2 } \right)^{n_{ij}^{u,d}},
\label{L}
\eeq
where $H^c=i\sigma_2 H^*$. Here we restrict our considerations to quarks  and  comment on the case of leptons at the end of sect.~\ref{sec_fla}. The coefficients $c_{ij}^{u,d} $ are numbers of order unity, while $n_{ij}^{u,d}$ are integers.

When the Higgs field develops a vacuum expectation value (VEV) $\vev{H}\equiv v \ll M$ (where $v=174$~GeV), the fermions attain hierarchically small masses, depending
on   $n_{ij}^{u,d}$. On the other hand, the coupling of the physical Higgs boson 
$h=\sqrt{2} {\rm Re }(H^0-v)$  to the fermions
with flavors $i,j$   increases by a factor $2 n_{ij}+1$ compared to that of the SM,
\beq
y_{ij}^{u,d}=   \left(2 n_{ij}^{u,d}+1\right) ~   ( y_{ij}^{u,d} )_{\rm SM}\;, \label{yij}
\eeq
where $( y_{ij}^{u,d} )_{\rm SM}=m^{u,d}_{ij}/(\sqrt{2}v)$.
For the top quark $n_{33}^u=0$, while suppression of  the bottom quark mass requires 
$n_{33}^d=1$. This fixes the expansion parameter
\beq
\epsilon \equiv \frac{v^2}{M^2} \simeq \frac{m_b}{m_t}  \simeq \frac{1}{60},
\label{eps}
\eeq
and therefore the new-physics scale $M$ must be about 1--2~TeV.
It is intriguing that  flavor physics  points independently towards the  
same scale that is favored by hierarchy-problem considerations.
Note that choosing $n_{33}^d>1$ would lead to an excessively low value of $M$.

The structure of Yukawa couplings in \eq{L} is reminiscent of the Froggatt-Nielsen approach~\cite{Froggatt:1978nt}.  However, there are three important differences. First of all, in the  Froggatt-Nielsen case, the new-physics scale is arbitrary and 
 often associated with some super-heavy mass. In our case, $M$ is necessarily around the TeV and this leads to a rich variety of phenomenological predictions that we will discuss in the following. 

The second difference is that our expansion parameter $\epsilon$ in \eq{eps} is smaller than the typical expansion parameter of the Froggatt-Nielsen approach, which is usually taken to be equal to the Cabibbo angle $\lambda =0.22$. This means that we will use the $\epsilon$ expansion to reproduce only the broad features of the fermion mass matrices and  allow the coefficients $c^{u,d}_{ij}$ to take values from about 1/5 to 5, in order to fit precisely   the mass and mixing parameters. 
In practice, we will choose appropriate values of $n_{ij}^{u,d}$ to obtain the hierarchy
\beq
\frac{m_t}{v}\sim V_{us} \sim {\cal O}(\epsilon^0), ~~~
\frac{m_{b,c}}{v}\sim V_{cb} \sim V_{ub}\sim {\cal O} (\epsilon^1),~~~
\frac{m_s}{v}\sim {\cal O}(\epsilon^2), ~~~
\frac{m_{u,d}}{v}\sim  {\cal O}(\epsilon^3).
\label{mas}
\eeq

The third important remark is that $H^\dagger H$ cannot carry any quantum number and thus  cannot play the role of
the Froggatt-Nielsen field. However, this becomes possible in the supersymmetric version of our approach, where
one  replaces $H^\dagger H$ with the gauge-invariant combination $H_u H_d$ of the two Higgs doublet superfields. The Yukawa couplings in the superpotential are holomorphic functions of this combination and can be expressed as 
\beq
Y_{ij}^{u,d}= c_{ij}^{u,d}  \left( { H_u H_d \over M^2 } \right)^{n_{ij}^{u,d}} .
\label{SUSY}
\eeq
Since the field combination $H_u H_d$ can carry a U(1) charge, it provides a direct analogy to the Froggatt-Nielsen field. The form of $Y_{ij}$
is dictated by U(1) charge conservation and with an appropriate charge assignment one can
reproduce the fermion masses and mixings. In this case, the Yukawa couplings have a ``factorizable''
form and 
\begin{equation}
n_{ij}^{u,d} = a_i + b_{j}^{u,d} \;, \label{ab}
\end{equation}
where $ a_i , b_{j}^{u,d}$ are related to the  U(1) charges of $q_{Li}$, $u_{Rj}$ and $d_{Rj}$, respectively. 
We also note that here the expansion parameter is $\epsilon = (v^2\sin2\beta)/(2M^2)$, which is smaller than that 
in the non-supersymmetric case for the same value of $M$, especially at large $\tan\beta$. However, large $\tan\beta$ also reduces the ratio between the top and bottom Yukawa couplings.

The absence of any quantum number associated with $H^\dagger H$ makes the form of the Lagrangian in \eq{L} potentially unstable under quantum corrections. Indeed, the power  $n_{ij}$ in the Yukawa interaction can be reduced by closing the $H^\dagger H$ lines through a Higgs propagator into a quadratically divergent one-loop diagram. Let us assume that all quadratic divergences associated with the Higgs are cut off by TeV-physics at the mass scale of the new states, which is equal to $g_FM$, where $g_F$ is the typical coupling of the unknown flavor dynamics.
The reduction  of $n_{ij}$ by one unit, obtained by the one-loop integration, carries a suppression factor of  about $g_F^2/(16\pi^2)$.  This effect is subleading to the hierarchy created by the $\epsilon$ expansion as long as
\beq
\frac{g_F^2}{16\pi^2}<\epsilon .
\label{loopexp}
\eeq
In the following, we will assume the validity of \eq{loopexp}, thus ensuring that the pattern of Yukawa couplings described by \eq{L} is rather  stable under radiative corrections. Note that \eq{loopexp} requires that flavor dynamics be rather weakly coupled ($g_F<4\pi \sqrt{\epsilon}\simeq 1$), and it gives an independent argument for fixing the flavor-dynamics mass scale $M$ in the TeV range ($M<4\pi v/g_F \simeq 2~{\rm TeV}/g_F$).

In general, the effective theory  also contains  Higgs-dependent fermionic kinetic terms of the form
\beq
{\cal L}_K=iZ_{ij}^{q}(H) {\bar q}_{Li} \slash{D} q_{Lj}+
 \Bigl( i\tilde Z_{ij}^{q}(H) {\bar q}_{Li} \slash{D} [ \alpha_{ij} (H)  q_{Lj} ] + {\rm h.c.} \Bigr) + ...,
\label{kin}
\eeq
where $Z^{q}$ is a  Hermitian matrix depending on  $H^\dagger H /M^2$,  $\tilde Z^{q}$
is a general matrix depending on $H$ and $  H^\dagger$, and $\alpha_{ij} (H)$ is a function of
$H$ and $  H^\dagger$ (e.g. it is either $H$ or $  H^\dagger$ to lowest order).  
The ellipses stand for higher derivative terms.
Similar expressions hold for $u_R$ and $d_R$ kinetic terms.

Consider the first term in \eq{kin}. 
When $H$ is replaced by its VEV, the kinetic terms can be made canonical by unitary rotations and field rescalings. These transformations affect the Yukawa matrices and, if the coefficients $Z_{ij}^{q,u,d}(v)$ contain non-trivial powers of $v/M$, they can contribute to generation of the  hierarchy in fermion masses and mixings. However, if quarks are elementary particles in the fundamental theory, then $Z^{q,u,d}=\identity + {\cal O} [(H^\dagger H /M^2)^n]$ (with $n>0$)  and the field rescaling gives only subdominant effects in $\epsilon$.
The terms in $Z^{q,u,d}$ depending on the Higgs field fluctuation
cannot be removed and  generate additional anomalous Higgs interactions. Since we are interested in processes in which the fermions are real particles in the external legs, we can use their equations of motion to reduce these interactions to the form of Yukawa couplings.
Therefore, for simplicity, in the following we will drop the effects of the kinetic terms in \eq{kin} and concentrate only on the Higgs-dependent Yukawa couplings.

The second term in  \eq{kin} yields corrections  to both the kinetic term and the $Z$- and $W$-couplings.
This is because  the gauge structure of $\slash{D} [ \alpha (H)  q_{L} ]$ is in general
different from that of $\skew0 \slash{D}   q_{L} $ and after electroweak symmetry breaking this produces a correction to the couplings between fermions and gauge bosons. Such a  correction in general splits into a flavor universal part, which affects electroweak precision physics, and a flavor violating part which is  constrained by flavor physics.
These effects can be studied only on a model-by-model basis. We also note that at 1-loop additional  
 dipole-type interactions  $\bar q_{Li} \sigma_{\mu \nu} 
q_{Rj} ~ F^{\mu \nu}$ are generated.

\section{Higgs-mediated flavor violation}
\label{sec_fla}

From \eq{yij} it immediately follows that the Higgs boson mediates FCNC at tree level.
The reason is that the fermion mass matrix and the matrix of Higgs--fermion couplings 
differ by a flavor dependent factor $2n_{ij}+1$, and thus they cannot be  diagonalized simultaneously.

Let us choose a field basis in which the quark mass matrices $m^{u,d}$ are diagonal and real, obtained through the bi-unitary transformation
\begin{equation}
Y^{u,d} (v)\rightarrow V_L^{ u,d \dagger} ~Y^{u,d}(v)~ V_R^{u,d} = \frac{m^{u,d}}{v} .
\end{equation}
Here $V_{L,R}^{u,d}$ are unitary matrices and the  CKM matrix is given by $V_{CKM}=V_L^{ u \dagger}V_L^{d}$.
In this basis, the interaction between a single Higgs boson and the fermionic current is given by
\beq
{\cal L}_h=-\frac{h}{\sqrt{2}}J_h,~~~~~J_h\equiv \frac{m_i^u}{v}{\bar u}_i u_i +2\left( G^u_{ij}~{\bar u}_{Li}u_{Rj} +{\rm h.c.}\right) + \left( u\leftrightarrow d \right) ,
\label{hint}
\eeq
\beq
G^{u,d}_{ij}\equiv \frac{m^{u,d}_k}{v} ~n^{u,d}_{lm}~{V^{u,d}_{L\; li}}^{*}~V^{u,d}_{L\; lk}~{V^{u,d}_{R\; mk}}^*~V^{u,d}_{R\; mj}.
\eeq
The coupling $G^{u,d}_{ij}$ describes the new interaction term, not present in the SM. Its expression becomes simpler under the factorization hypothesis of \eq{ab},
\beq
G^{u,d}_{ij}=A^{u,d}_{ij}\frac{m^{u,d}_j}{v}+ \frac{m^{u,d}_i}{v}B^{u,d}_{ij},
~~~~~~~A^{u,d}\equiv V_L^{u,d\dagger} aV_L^{u,d},~ B^{u,d}\equiv V_R^{u,d\dagger} b^{u,d}V_R^{u,d}.
\label{gud}
\eeq
The information about flavor violation is contained in the Hermitian matrices $A^{u,d}$ and $B^{u,d}$. To simplify our discussion, we will focus on factorizable Yukawa matrices, but analogous considerations can be made for the general case.

By integrating out the Higgs boson at tree level, we obtain the four-fermion effective interaction Lagrangian
\beq
{\cal L}_{4f}=\frac{J_h^2}{4m_h^2},
\label{fourf}
\eeq
where $m_h$ is the Higgs boson mass. The Lagrangian in \eq{L} also contains multi-Higgs interactions with quarks of the form $m_{ij}{\bar q}_iq_j(h/v)^{p+1}$, where $p\le 2n_{ij}$ and $m_{ij}$ is the quark mass matrix in the current eigenbasis. By integrating out the Higgs bosons at $p$ loop order, these interactions also generate terms in the four-fermion effective Lagrangian. The ratio between the coefficients of the $p$-loop contribution and the tree-level contribution is of the order of
\beq
\left( \frac{g_F M}{4\pi v}\right)^{2(p-1)} \left( \frac{m_h}{4\pi v}\right)^2,
\eeq
where we have cut off power-divergent integrals at the mass scale of the new states $g_FM$.
Using \eq{loopexp} and the requirement of a perturbative Higgs quartic coupling ($m_h<4\pi v$), we observe that loop contributions are always subleading with respect to the tree-level effect given in \eq{fourf}.  

The flavor-violating effects can be easily extracted from \eq{fourf}. For instance, the $\Delta S=2$ interaction is given by
\beq
{\cal L}_{\Delta S=2} = \frac{m_s^2}{v^2m_h^2} \left( A^d_{12}~{\bar d}_Ls_R+B^d_{12}~{\bar d}_Rs_L\right)^2.
\eeq
This gives a contribution to the mass difference of the neutral kaons 
\beq
\frac{\Delta m_K}{m_K}\simeq \frac{5f_K^2m_K^2}{12v^2m_h^2}\left( {A^{d}_{12}}^2+{B^{d}_{12}}^2-\frac{12}{5}A^{d}_{12}B^{d}_{12}\right) .
\eeq
Analogous results can be obtained for the mass differences in the $B^0$ and $D^0$ systems. Requiring  that the new contributions do not exceed the experimental bounds, we obtain the following constraints
\bea
\sqrt{\left| {A^{d}_{12}}^2+{B^{d}_{12}}^2-\frac{12}{5}A^{d}_{12}B^{d}_{12}\right|} &<& 6\times 10^{-2}~\frac{m_h}{200~{\rm GeV}} \label{mk}\\
\sqrt{\left| {A^{d}_{13}}^2+{B^{d}_{13}}^2-\frac{14}{5}A^{d}_{13}B^{d}_{13}\right|} &<& 2\times 10^{-2}~\frac{m_h}{200~{\rm GeV}} \label{mbd}\\
\sqrt{\left| {A^{d}_{23}}^2+{B^{d}_{23}}^2-\frac{14}{5}A^{d}_{23}B^{d}_{23}\right|} &<& 7\times 10^{-2}~\frac{m_h}{200~{\rm GeV}} \label{mbs}\\ 
\sqrt{\left| {A^{u}_{12}}^2+{B^{u}_{12}}^2-\frac{14}{5}A^{u}_{12}B^{u}_{12}\right|} &<& 2\times 10^{-2}~\frac{m_h}{200~{\rm GeV}} . \label{md}
\eea
These constraints are satisfied as long as $A^{u,d}_{12}$, $B^{u,d}_{12}$, $A^d_{13,23}$, $B^d_{13,23}$ are all ${\cal O}(\epsilon )$ or smaller. From the definition of the matrices $A^{u,d}$ and $B^{u,d}$ in \eq{gud} it is easy to see that their off-diagonal elements $ij$ are suppressed by powers of $\epsilon$ either if the corresponding mixing angle in the rotation matrices is suppressed ($V_{L,R\; ij}\sim \epsilon$) or if the corresponding $a$ or $b$ coefficients are universal ($a_i=a_j$ or $b_i=b_j$). As we will show in the following, for realistic assignments that reproduce the quark mass pattern, this is always the case, and the off-diagonal elements of the matrices  $A^{u,d}$ and $B^{u,d}$ are indeed ${\cal O} (\epsilon )$ or smaller. 

Constraints from $\Delta F=1$ dipole processes like $b\rightarrow s \gamma$ are less restrictive,
because the new contribution is suppressed by a loop factor and  three Yukawa couplings\footnote{Similarly, Higgs-induced EDMs are small since
they are also suppressed by a loop factor and  three Yukawa couplings.}.
On the other hand, the CP-violating observable $\epsilon_K$ imposes a severe constraint on the new complex  phases present in the effective theory,
\beq
\sqrt{{\rm Im}~\left( {A^{d}_{12}}^2+{B^{d}_{12}}^2-\frac{12}{5}A^{d}_{12}B^{d}_{12}\right)} < 4\times 10^{-3}~\frac{m_h}{200~{\rm GeV}} .
\label{epl}
\eeq
The Yukawa structures consistent with the size of the Cabibbo angle typically predict that either $A^d_{12}$ or 
$B^d_{12}$ is ${\cal O}(\epsilon)$ and therefore the constraint in \eq{epl} is not naturally satisfied. However, in our approach, some of the $c^d_{ij}$ coefficients in \eq{L} have to be somewhat smaller than one, and therefore it is possible to arrange their values 
such that the constraint in \eq{epl} is satisfied. Since in the limit $c_{21}^d \ll 1$, the CP phases can
be rotated away from the upper left 2$\times$2 block of $Y^d$, choosing $|c^d_{21}|\le 0.1$ allows us to satisfy the constraint from $\epsilon_K$, without altering the quality of the fit to the quark masses and mixings.  Alternatively, we can assume that the CP-violating phase entering \eq{epl} (not related to the CKM phase) is about 0.1, rather then 1. This would be sufficient to satisfy the constraint. Finally, we mention that 
the CP-violating observable 
$\epsilon_K^\prime$ imposes only a weaker constraint on Im$A_{12}^d$ and Im$B_{12}^d$.

Let us  now give an example of a texture that leads to the mass and mixing pattern of \eq{mas}.
Choosing
\begin{equation}
a=(1,1,0)~,~b^d=(2,1,1)~,~b^u=(2,0,0)\;,
\label{assab}
\end{equation}
we obtain\footnote{Equation~(\ref{assab}) actually describes the unique factorizable and hierarchical solution leading to  \eq{mas}. If we modify the assumption of \eq{mas}, taking $V_{us} \sim V_{cb}\sim O(\epsilon^1)$ and $V_{ub}\sim O(\epsilon^2)$, there is a unique solution with $a=(2,1,0)~,~b^d=(1,1,1)~,~b^u=(1,0,0)$ leading to
\begin{equation}
A^{u,d}= \left(
\begin{matrix}
2 & \epsilon^1 & \epsilon^2 \\
\epsilon^1 & 1 & \epsilon^1 \\
\epsilon^2 & \epsilon^1 & \epsilon^2 \\
\end{matrix}
\right)~,~
B^u = \left(
\begin{matrix}
1 & \epsilon^1 & \epsilon^1 \\
\epsilon^1 & \epsilon^2 & \epsilon^2 \\
\epsilon^1 & \epsilon^2 & \epsilon^2 \\
\end{matrix}
\right)~,
\nonumber
\end{equation}
and $B^d$ equal to the identity.
Of course, more solutions exist if we drop the factorization hypothesis of \eq{ab}.}
\begin{equation}
Y^d\sim \left(
\begin{matrix}
\epsilon^3 & \epsilon^2 & \epsilon^2 \\
\epsilon^3 & \epsilon^2 & \epsilon^2 \\
\epsilon^2 & \epsilon^1 & \epsilon^1 \\
\end{matrix}
\right)~,~
Y^u\sim \left(
\begin{matrix}
\epsilon^3 & \epsilon^1 & \epsilon^1 \\
\epsilon^3 & \epsilon^1 & \epsilon^1 \\
\epsilon^2 & \epsilon^0 & \epsilon^0 \\
\end{matrix}
\right)~. \label{texture}
\end{equation}
The corresponding matrices $A^{u,d}$ and $B^{u,d}$, at leading order in $\epsilon$, are given by 
\begin{equation}
A^{u,d}= \left(
\begin{matrix}
1 & \epsilon^2 & \epsilon^1 \\
\epsilon^2 & 1 & \epsilon^1 \\
\epsilon^1 & \epsilon^1 & \epsilon^2 \\
\end{matrix}
\right)~,~
B^d = \left(
\begin{matrix}
2 & \epsilon^1 & \epsilon^1 \\
\epsilon^1 & 1 & \epsilon^2 \\
\epsilon^1 & \epsilon^2 & 1 \\
\end{matrix}
\right)~,~
B^u = \left(
\begin{matrix}
2 & \epsilon^2 & \epsilon^2 \\
\epsilon^2 & \epsilon^4 & \epsilon^4 \\
\epsilon^2 & \epsilon^4 & \epsilon^4 \\
\end{matrix}
\right)~.
\label{abmat}
\end{equation}
Since the off-diagonal elements of $A^{u,d}$ and $B^{u,d}$ are $O(\epsilon)$ or smaller, all $\Delta F=2$ constraints are satisfied. The Higgs-mediated contribution to $\Delta m_{B_d}$ is significant, nearly saturating the experimental value, 
as seen from \eq{mbd}. On the other hand, since both $A^u_{12}$ and $B^u_{12}$ are $O(\epsilon^2)$, the new contribution to $\Delta m_{D}$ is rather small. 
We have checked that the texture defined by \eq{assab} can reproduce the known values of quark masses and mixing, taking all coefficients $c_{ij}^{u,d}$ in the range 1/5 to 5. 

The extension of the Higgs-dependent Yukawa couplings hypothesis to the charged lepton sector is straightforward. The Higgs boson mediates interactions that violate lepton flavor and contribute to rare $\mu$ and $\tau$ decays and $\mu$--$e$ conversion processes. A simple estimate gives
\bea
BR(\mu \to e\gamma ) \sim \frac{\alpha}{4\pi} \left( \frac{m_\mu^2A^\ell_{12}}{m_h^2}\right)^2 &=&\left( \frac{200~{\rm GeV}}{m_h}\right)^4 \left( \frac{A^\ell_{12}}{1/60}\right)^2~1\times 10^{-20}\\
BR(\mu \to eee ) \sim \left( \frac{m_\mu m_eA^\ell_{12}}{m_h^2}\right)^2 &=&\left( \frac{200~{\rm GeV}}{m_h}\right)^4 \left( \frac{A^\ell_{12}}{1/60}\right)^2~5\times 10^{-22}\\
BR(\tau \to \mu\gamma ) \sim \frac{\alpha}{4\pi} \left( \frac{m_\tau^2A^\ell_{23}}{m_h^2}\right)^2 &=&\left( \frac{200~{\rm GeV}}{m_h}\right)^4 \left( \frac{A^\ell_{23}}{1/60}\right)^2~1\times 10^{-15}\\
BR(\tau \to \mu\mu\mu ) \sim \left( \frac{m_\tau m_\mu A^\ell_{23}}{m_h^2}\right)^2 &=&\left( \frac{200~{\rm GeV}}{m_h}\right)^4 \left( \frac{A^\ell_{23}}{1/60}\right)^2~6\times 10^{-15}.
\eea 
The off-diagonal elements of the matrices $A^\ell$ and $B^\ell$ are not related to CKM angles. However, even if they have the same form as the corresponding elements in the down-quark sector and  are of order $\epsilon$, the predicted rates for lepton flavor violation satisfy the present experimental constraints and  are too small to give a detectable signal in planned experiments. 

We note that,  with Higgs-dependent Yukawa couplings, CP violation is
already possible in the system of two quarks, say the top and charm quarks.
The analog of the Jarlskog invariant that controls CP violation in Higgs interactions
is Im[Tr$(m y^\dagger)^2$]. Unlike in the SM, it is not suppressed by a product
of light quark masses. This opens up the possibility of EW baryogenesis which
is to be studied elsewhere.

\section{Higgs physics}
\label{sec_hig}

One of the most striking features of our approach is that the Higgs branching ratios are drastically modified with respect to the SM predictions. The Higgs couplings to the weak gauge bosons and to the top quark (and consequently to gluons\footnote{The modification of the Higgs-bottom coupling can affect the Higgs-gluon coupling, but the effect on the gluon-fusion rate is at most at the 10\% level.} and photons) remain the same as those in the SM, but 
\beq
\frac{\Gamma \left( h\to b\bar b\right)}{\Gamma \left( h\to b\bar b\right)_{SM}}=
\frac{\Gamma \left( h\to c\bar c\right)}{\Gamma \left( h\to c\bar c\right)_{SM}}=
\frac{\Gamma \left( h\to  \tau^+\tau^- \right)}{\Gamma \left( h\to \tau^+\tau^- \right)_{SM}}=9,
~~~~\frac{\Gamma \left( h\to  \mu^+\mu^- \right)}{\Gamma \left( h\to \mu^+\mu^- \right)_{SM}}=25.
\eeq
The Higgs couplings to light quarks are  enhanced even further, but they do not play any significant role in Higgs phenomenology. The enhancement of the coupling to $\mu$ can be very important in view of a possible muon collider, since the Higgs production rate is predicted to be a factor of 25 larger than that in the SM.

\begin{figure}[t!]
\begin{center}
\includegraphics[width=14cm]{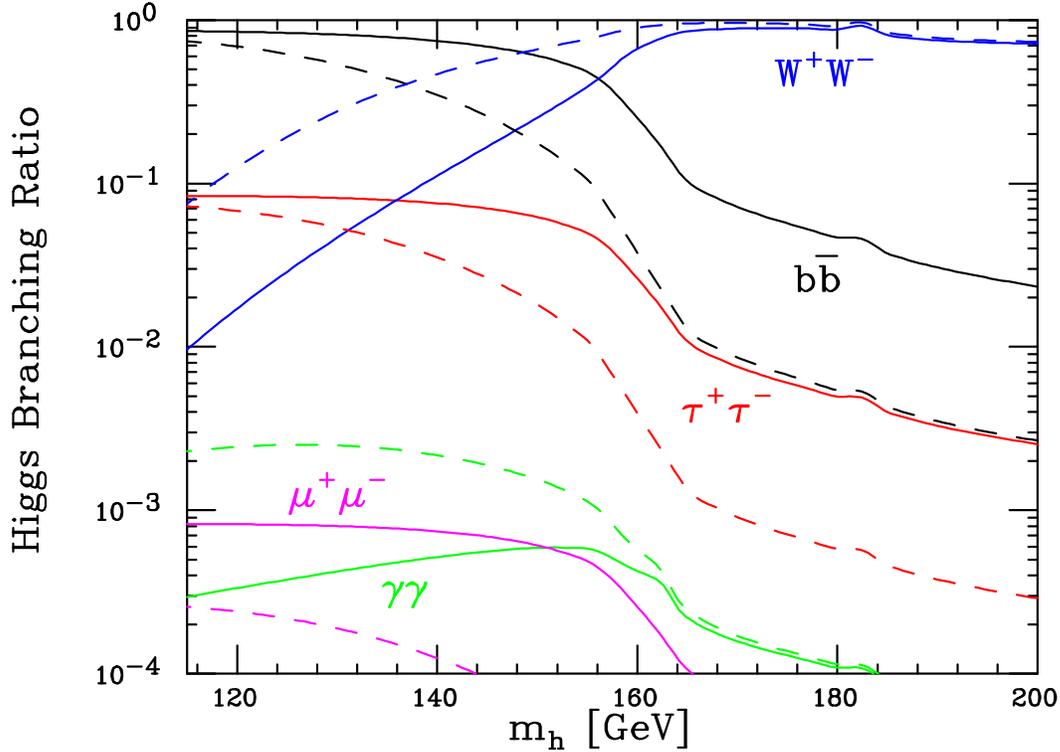}
\end{center}
\caption{
The various color lines show the Higgs branching ratios for different decay modes, with solid lines referring to the case of Higgs-dependent Yukawa couplings and dashed lines to the SM. 
 }
 \label{fig1}
\end{figure}

\begin{figure}[t!]
\begin{center}
\includegraphics[width=14cm]{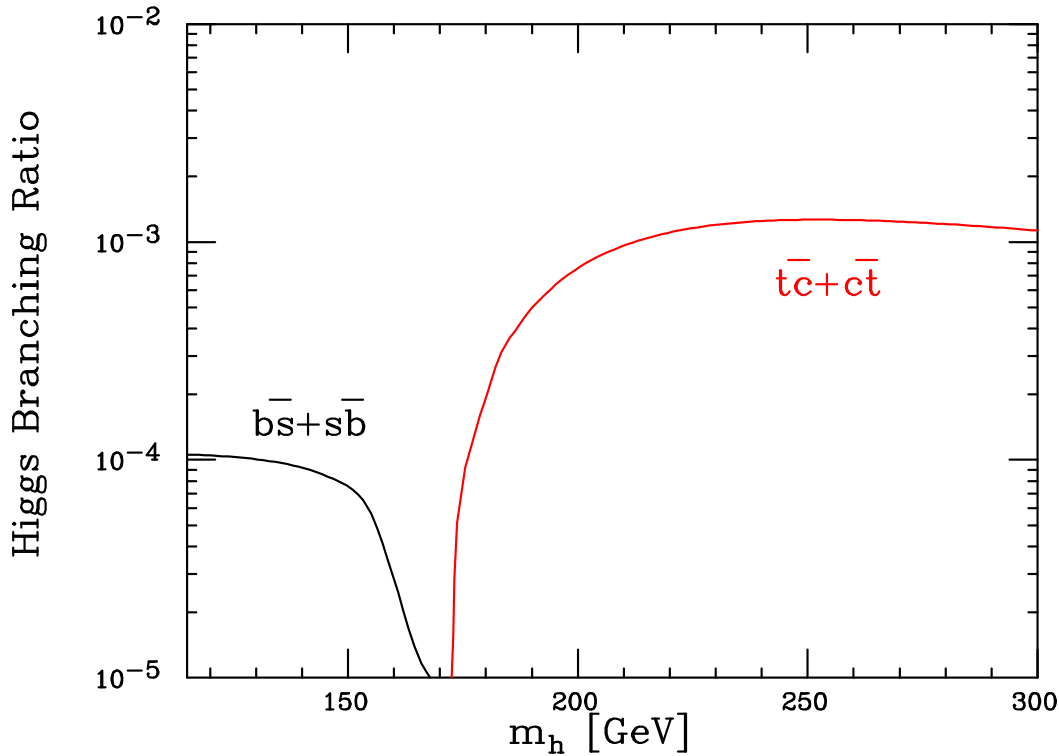}
\end{center}
\caption{
The Higgs branching ratios for flavor-violating decay modes, in the case of Higgs-dependent Yukawa couplings, taking  $A^{u,d}_{23} =\epsilon  =m_b/m_t$ and $B^{u,d}_{23} =0$.
 }
 \label{fig2}
\end{figure}

All main Higgs production processes at the LHC such as the gluon fusion, weak-boson fusion, Higgs-strahlung from the top or gauge boson are not affected, but the novelty lies in the Higgs decay. In fig.~\ref{fig1} we show the prediction for the Higgs branching ratios in the most important channels. At low $m_h$, there is an increase of $BR( h\to b\bar b )$ and $BR( h\to  \tau^+\tau^- )$ with respect to those in the SM, while, more importantly, there is also a significant reduction of  $BR( h\to \gamma \gamma)$. Actually, the Higgs decay rate into muons becomes even larger than the one into photons, although its branching ratio remains smaller than $10^{-3}$. At intermediate values of $m_h$, the main effect is an increase of the decay rate into fermions compared  to that into $WW$ and $ZZ$. As a result, $h\to WW$ becomes the leading decay mode only for $m_h>156$~GeV, while in the SM this happens for $m_h>136$~GeV.

Another peculiarity of our scenario concerns  flavor-violating Higgs decay modes. The new Higgs interactions can be read off from \eq{hint},
\beq
{\cal L}_h=-\frac{\sqrt{2}h}v ~{\bar u} \big[ m^u\left( A^u P_L +B^uP_R\right) + \left( A^u P_R +B^uP_L\right) m^u \big] u + \left( u \leftrightarrow d \right) ,
\eeq
where $P_{R,L}=(1\pm \gamma_5)/2$ are the chiral projectors and $m^{u,d}$ are the diagonal quark mass matrices. From this we obtain
\beq
\frac{\Gamma \left( h\to b\bar s +\bar b s \right)}{\Gamma \left( h\to b\bar b\right)_{SM}}=
4\left( \left| A^d_{23}\right|^2 +  \left| B^d_{23}\right|^2 \right) ,
\label{hbs}
\eeq
\beq
 \frac{\Gamma \left( h\to t\bar c +\bar t c \right)}{\Gamma \left( h\to b\bar b\right)_{SM}}=
4\left( \left| A^u_{23}\right|^2 +  \left| B^u_{23}\right|^2 \right) \frac{m_t^2}{m_b^2} \left( 1- \frac{m_t^2}{m_h^2}\right)^2 .
\label{htc}
\eeq
For $A^{u,d}_{23} =O(\epsilon )$, as given  in \eq{abmat}, the ratio in \eq{hbs} is about $10^{-3}$.
In \eq{htc}, the powers of $\epsilon$ from $A^u_{23}$ are exactly compensated by the factor $m_t^2/m_b^2$ and the flavor-violating Higgs decay into the top quark, whenever kinematically allowed, has a rate of the order of the SM width for $h\to b\bar b$.  We note however that 
 the process $h\to tc$ is allowed only in the region of $m_h$ where the dominant Higgs decay channel is $h\to WW$.
In fig.~\ref{fig2} we show our prediction for the branching ratios of the flavor-violating Higgs modes, taking  $A^{u,d}_{23} =\epsilon  =m_b/m_t$ and $B^{u,d}_{23} =0$. In the range $200~{\rm GeV}\simlt m_h \simlt 300~{\rm GeV}$, $BR( h\to t\bar c +\bar t c )$ is typically of the order of $10^{-3}$. For comparison, the corresponding branching ratio in the SM is of order $10^{-13}$ and that in two-Higgs doublet models is always
less than $10^{-4}$ (see {\it e.g.} ref.~\cite{Baum:2008qm}). 

If the Higgs is lighter than the top quark, the interesting flavor-violating process is $t\to hc$. Its branching ratio is
\beq
BR\left( t\to hc\right) =\frac{2 \left( \left| A^u_{23}\right|^2 +  \left| B^u_{23}\right|^2 \right) \left( 1- \frac{m_h^2}{m_t^2}\right)^2}{\left( 1- \frac{3m_W^4}{m_t^4}+\frac{2m_W^6}{m_t^6}\right)}.
\eeq
For $A^{u,d}_{23} =O(\epsilon )$ and  a light Higgs, $BR\left( t\to hc\right)$ is about $10^{-3}$, which is well within the reach of the LHC, since experiments are expected to probe values down  to $5\times 10^{-5}$~\cite{branco}. 
We note that the corresponding SM prediction is $6\times 10^{-15}$, while type I and II two-Higgs doublet models give $BR\left( t\to hc\right) < 10^{-5}$ (see {\it e.g.} ref.~\cite{Baum:2008qm}).

\section{Example of TeV completion}
\label{sec_mod}

In sect.~\ref{sec_fla} we have investigated  flavor violation generated by Higgs exchange within the effective theory. However, since the mass $M$ is around the TeV scale, higher-dimensional operators involving quarks and leptons, obtained from integrating out the heavy modes, can  potentially be a dangerous source of  FCNC. This issue cannot be addressed without specifying a  particular model of flavor dynamics at the scale $M$. In this section we will present an example of a completion of the effective theory beyond the scale $M$, which does not lead to excessive flavor violation. Our example is not meant to describe a fully realistic theory of flavor, but only to illustrate how Higgs-dependent Yukawa couplings can be generated through couplings of quarks and leptons to some heavy fields.

Consider an extension of the SM with some heavy vectorlike Dirac fermions $S^d, R^d$ having the gauge quantum numbers of  the down-quark chiral components, i.e. $S^d \sim q_L$, $R^d \sim d_R$, with the interaction Lagrangian
\begin{eqnarray}
-{\cal L} &=& \left[ \bar q_L \lambda_0^d d_R+  \bar q_L \lambda_1^d R^d +
\bar S^d \lambda_2^d d_R
+  \bar S^d \left( {\lambda_3^{d}}^\dagger P_L +\lambda_4^d P_R\right) R^d\right] H +{\rm h.c.} \nonumber\\
&+& \bar S^d m_S^d S^d  + \bar R^d m_R^d R^d \;.
\label{lagrhe}
\end{eqnarray}
Here $\lambda_{0-4}^{d}$ and $m_S,m_R$ are 
matrices in generation space and we have suppressed the flavor indices. 
We have also
made the kinetic terms canonical and eliminated, by an appropriate basis 
transformation, possible mass terms which mix the light and heavy generations. We also introduce heavy fields $S^{u}$ and $R^{u}$ for the up-quarks  in an analogous way. For simplicity, we assume that the heavy fields from the up quark sector have negligible interactions
with the down quark sector.  
The couplings $\lambda_i^d$ and the mass scale $m_{S,R}/\lambda_i^d$ play the role of the parameters $g_F$ and $M$ that we have introduced in the effective theory in sect.~\ref{sec_eff}.

Using their equations of motion, we integrate out the heavy states from \eq{lagrhe} and 
 obtain modified kinetic terms for the quarks as well as  the following Yukawa couplings 
\bea
Y^d&=& \lambda_0^d + \tilde \lambda_1^d \tilde \lambda_3^d \Bigl(1-  \tilde \lambda_4^d \tilde \lambda_3^d\Bigr)^{-1} 
\lambda_2^d \nonumber \\
&=& \lambda_0^d + \tilde \lambda_1^d \tilde \lambda_3^d \sum_{n=0}^\infty \left( \tilde \lambda_4^d \tilde \lambda_3^d \right)^n \lambda_2^d 
\;, \label{heavyY}
\eea
where
\begin{equation}
\tilde \lambda_{1,4}^d = \lambda_{1,4}^d {1\over m_R^d} H ~,~  
\tilde \lambda_{3}^d = \lambda_{3}^d {1\over m_S^d} H^\dagger \;,
\label{deftil}
\end{equation}
and analogously for the up-quark sector. 
Expanding the result in powers of 
$\tilde \lambda_4^d \tilde \lambda_3^d$, we obtain the desired Higgs--dependent
Yukawa couplings as a series in $\epsilon$.

Although generically the Yukawa couplings in \eq{heavyY} do not have the
factorizable form of \eq{ab},
with an appropriate choice of the flavor matrices $\lambda_i$,
one can reproduce the texture of \eq{texture}. We find that in order  
to generate the texture in \eq{texture}, while avoiding excessive FCNC, one 
has  to introduce 4  generations of fermions $S$ and $R$.
The simplest   solution is to take  equal masses for all $S$ 
and $R$, flavor-conserving and universal couplings between the light 
and heavy fermions, while having all flavor violation reside in the 
interactions among the heavy fermions. For instance, we can choose
\beq
m_{S,R}= M ~ \identity_{4\times 4} ~,~ \lambda_{1}^{u,d}= \identity_{3
\times 4}~,~
\lambda_{2}^{u,d}= \identity_{4\times 3}~,
\label{univer}
\eeq
\begin{eqnarray}
&& \lambda_3^d= \left(
\begin{matrix}
0&0&0&1\\
0&0&0&1\\
0&1&1&0\\
1&0&0&0\\
\end{matrix}
\right)~,~
\lambda_4^d= \left(
\begin{matrix}
0&0&0&0\\
0&0&0&0\\
0&0&0&1\\
0&0&1&0\\
\end{matrix}
\right)~,~ \nonumber\\
&& \lambda_3^u= \left(
\begin{matrix}
0&1&1&0\\
0&1&1&0\\
0&0&0&1\\
1&0&0&0\\
\end{matrix}
\right)~,~
\lambda_4^u= \left(
\begin{matrix}
0&0&0&0\\
0&0&1&0\\
0&0&0&0\\
0&0&0&1\\
\end{matrix}
\right)~,~
\label{lamtex}
\end{eqnarray}
where $M\sim 1-2$ TeV,
$ (\identity_{n\times m})_{ij}$ is an $n\times m$  matrix with 
elements 1 if $i=j$ and 0
otherwise, and  ones and zeros in $\lambda_{3,4}^{u,d}$  are 
understood in the ``texture'' sense. The 
contribution to the quark wavefunctions from integrating out the 
heavy fields affects the Yukawa couplings at order   $
\epsilon^2$,  which, for our purposes, can be neglected.
Finally, the couplings of the light quarks among themselves are
\begin{equation}
\lambda_0^d=0~,~ \lambda_0^u=\left(
\begin{matrix}
0&0&0\\
0&0&0\\
0&1&1
\end{matrix}
\right)~.
\label{l0}
\end{equation}

\begin{figure}[t!]
\centerline{\includegraphics{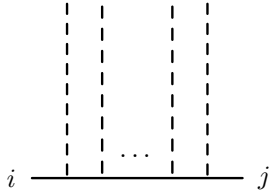}}
\caption{Fermion flavor violation via Higgs line insertions. The external legs of the fermion line represent quarks of 
generations $i$ and $j$, while the internal propagators represent the heavy fermions $R$ and $S$.
\label{FV}}
\end{figure}

Let us consider in more detail how FCNC are generated in this model.
An important ingredient that allows us to  suppress FCNC is the universality 
requirement in \eq{univer}, while \eq{lamtex} represents a particular 
choice of $\lambda_i$ which produces  the desired Yukawa structure.
Flavor violating operators are best studied in the weak basis
rather than the mass eigenstate basis. Any Feynman diagram 
contributing to flavor violation
is built out of  fermion lines with insertions of Higgs  (and 
possibly gauge boson) lines, see fig.~\ref{FV}. The Higgs lines  
either correspond to a Higgs VEV
insertion or they are closed in loops (among themselves or against 
another fermion line). The resulting flavor structure
can be studied order by order in the number of Higgs lines.
Every two Higgs lines 
correspond to a factor of $\epsilon$ regardless of whether they 
represent the Higgs VEV insertions or they are closed in a loop
(since $\epsilon$ is numerically  close to the loop factor). 
Keeping the number of Higgs lines  fixed, one sums over all possible
$\lambda_i$ matrices at the Higgs vertices.

The string of Higgs insertions in fig.~\ref{FV} always starts and 
ends with the matrices $\lambda_{1}$ or  $\lambda_{2}$ (or their 
conjugates), because we are interested in diagrams with external light 
quark lines. Since we require  $\lambda_{1,2}$ to be unit matrices,
non-trivial flavor structures can only arise  if
more than two Higgs lines are present.
  Consider the down sector. For 3 Higgs lines, the only possible
structures are
\begin{equation}
\bar d_{L_i} ~ \bigl( ~\lambda^d_1 \lambda^d_3 \lambda^d_2,~   \lambda^d_1 \lambda^{d\dagger}_4 \lambda^d_2 ~ \bigr)_{ij} ~d_{R_j}
  \;,
\end{equation}
which could potentially generate processes like the $b\rightarrow s \gamma$
transition when a photon line is attached and two Higgs lines are closed in a loop. 
However, $ \lambda^d_1 \lambda^{d\dagger}_4 \lambda^d_2 $ is zero for our
$\lambda_i$. Furthermore, $\lambda^d_1 \lambda^d_3 \lambda^d_2$ 
     is proportional to the mass matrix
for the down quarks (at order
$\epsilon$) in \eq{heavyY}, and thus the transformations that 
diagonalize the mass
matrix will also diagonalize flavor along this fermion line. Thus no 
FCNC are generated at this level.

The above cancellation is due to our  universality requirement in \eq{univer}, 
which ensures that the matrices $\tilde \lambda_i$ defined in \eq{deftil} are 
proportional to the corresponding matrices $\lambda_i$. Effectively, there is an operative GIM mechanism which 
ensures that  flavor violation is exactly rotated away, once we go 
to the mass eigenbasis. Loop corrections spoil the mass degeneracy of 
the heavy fermions, and we expect off-diagonal entries for the 
matrices $m_{S,R}$ of typical size $\left( \delta {m_{S,R}}\right)_{ij} =\left({m_{S,R}}\right)_
{ii}\lambda^2/(16\pi^2)$. These effects are taken into account by a 5-
Higgs insertion in the fermionic line, in which two Higgs vertices 
are closed in a loop by a Higgs propagator. The resulting 
contribution is sufficiently small and obeys the 
 experimental  limits on FCNC.

\begin{figure}[t!]
\centerline{\includegraphics{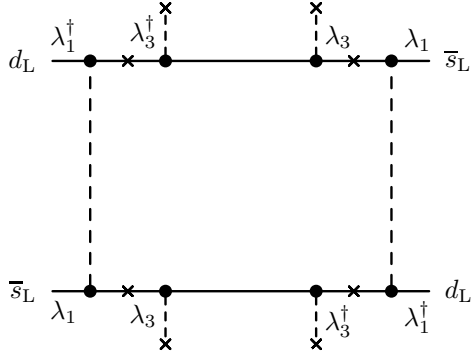}}
\caption{Example of a heavy quark contribution to the $K -\bar K$ mixing.
Crosses indicate either a heavy quark mass or a Higgs VEV insertion.
\label{Kaon}}
\end{figure}

In the case of 4 Higgs lines, the transition along the fermion line
is chirality conserving and a number of flavor objects appear.
Let us consider the left-left
transitions only, since the analysis for the right-right transitions is completely analogous. 
The possible structures along the fermionic string are
\begin{equation}
\bar d_{L_i}~ \left[ \left( 
  \lambda_1^d \lambda_3^d \lambda_4^d  \lambda_1^{d \dagger} +{\rm h.c.}\right),~\lambda_1^d \lambda_3^d \lambda_3^{d\dagger}  \lambda_1^{d
\dagger} ,~
\lambda_1^d \lambda_4^{d\dagger} \lambda_4^d  \lambda_1^{d\dagger} ~
    \right]_{ij} ~ d_{L_j} \;.
\label{struc4}
\end{equation}
All of these combinations but the last one also  contribute at tree 
level
(order $\epsilon^2$)  to  the  quark
kinetic terms $\bar d_{L_i} \not \! \partial d_{L_j}$.
The term  $\lambda_1^d \lambda_4^{d\dagger} \lambda_4^d  \lambda_1^{d\dagger}$
is generated when all heavy quark propagators preserve chirality
and thus leads to a higher-derivative operator. Although it does not 
affect  the quark kinetic terms, it contributes, for instance,
to the $\Delta F=2$ box diagram obtained by coupling the Higgs 
propagators to another identical fermion line. This creates a 
mismatch between the flavor structure
of the kinetic terms and that of the four-fermion operators. The 
transformations that diagonalize the kinetic terms do not then
diagonalize  flavor along the fermion line of the box diagram, leading
to FCNC. In this case, the universality requirement in \eq{univer} is 
not sufficient to guarantee a complete GIM mechanism. Thus FCNC 
appear at order $\epsilon^2$ times a loop factor.

The most dangerous  FCNC operator induced  is $(\bar s_L \gamma_\mu 
d_L)^2$.
An estimate of the diagram in fig.~\ref{Kaon} yields     
\begin{equation}
{\cal H}_{\rm eff} \simeq   {\epsilon^2  \over 640 \pi^2 M^2      }    (\bar s_L \gamma_\mu 
d_L)^2 \;.
\end{equation}
This induces $\Delta m_K$ of order $10^{-16}$ GeV which is below the experimental limit.
Other meson mixing constraints are weaker.
On the other hand $\epsilon_K$ is quite restrictive in general and arbitrary CP phases typically
overproduce it by an order of magnitude.
This constraint, however,
depends on  how CP violating phases enter
the flavor structures. For example, if CP violation is mostly due to 
the
up quark sector interactions, the  $\epsilon_K$ constraint is not
significant.
We also note that the heavy quark contribution to the real and imaginary parts of the $K-\bar K$ mixing
can be suppressed further for other choices of $\lambda_{3,4}^d$. In particular,  
we have found  $5\times 5$ textures that  induce $K-\bar K$ mixing only at order $\epsilon^3$.

Interactions with the heavy quarks also induce non-trivial kinetic structures of the form of the second term in
\eq{kin}.
To lowest order these terms are
\begin{equation}
{(\lambda_1^d \lambda_1^{d\dagger})_{ij} \over M^2} 
\overline{(H^\dagger q_{Li})} i \slash{D}(H^\dagger q_{Lj}) + 
{(\lambda_1^u \lambda_1^{u\dagger})_{ij} \over M^2} 
\overline{(H  q_{Li})} i\slash{D}(H  q_{Lj}) \;,
\end{equation}
and analogous terms are generated for $u_R$ and $d_R$. Here $H q_L \equiv   \epsilon^{ab}  H_a q_{Lb} $ with 
$a,b$ being the SU(2) indices. 
Since $\slash{D}(H^\dagger q_{Lj})$ and 
$\slash{D}(H  q_{Lj})$ have a different gauge structure compared to that of $\slash{D} q_{Lj}$,
we find a modification of
the $Z$-fermion  couplings. 
After canonically redefining the fields, the tree-level couplings $g_{L,R}$ of the left and right-handed quarks to the $Z$ are given by
\bea
g_L=I_3 -\sin^2\theta_W ~Q +\delta g_L ~~~~&&~~~ \delta g_L= -I_3 \frac{ \lambda_1^{u,d} \lambda_1^{u,d\dagger}v^2}{M^2}, \nonumber \\
g_R= -\sin^2\theta_W ~Q +\delta g_R ~~~~&&~~~ \delta g_R= I_3\frac{ \lambda_2^{u,d\dagger} \lambda_2^{u,d}v^2}{M^2} .
\label{glgr}
\eea
Here $I_3$ is the third isospin component ($I_3=\pm 1/2$ for up and down quarks, respectively) and $Q$ is the fermion electric charge. 
For universal $\lambda_{1,2}^{u,d}$ and $m_{R,S}$, as in \eq{univer},
the correction affects only $g_A \equiv g_L-g_R$ and not $g_V \equiv g_L+g_R$, and we find $\delta g_A = O(\epsilon)$. For $\epsilon  \sim 10^{-2}$, which is appropriate to reproduce the hierarchical pattern of \eq{mas},  $\delta g_A$ is not inconsistent with experimental data on $g_A$ for charm and bottom quarks, since the 2$\sigma$ errors are $\Delta g_A^c=1.06\times 10^{-2}$ and $\Delta g_A^b=1.02\times 10^{-2}$~\cite{lepewwg}.
Actually, if we simultaneously include the corrections to all quark couplings, we find that the strongest constraint is imposed by $R_b=0.21629\pm 0.00066$~\cite{lepewwg} which allows for a 0.6\% deviation at the 2$\sigma$ level. The corrections from \eq{glgr} induce $\delta R_b/R_b=0.28\epsilon$, which is $0.5\%$ for $\epsilon =1/60$.
Note that there is freedom in the model to  reduce the contributions to $\delta g_{L,R}$ while keeping the quark masses fixed, by decreasing the values of $\lambda_{1,2,4}$ and increasing $\lambda_{3}$. This rescaling  is limited by the requirement of maintaining perturbative couplings. Finally, similar corrections to the $W$-fermion couplings are relatively weakly constrained and do not lead to further bounds.

The situation is more problematic when we try to extend the model to leptons, since $g_A^\ell$ has been measured with the  precision of $10^{-3}$ at 2-$\sigma$~\cite{lepewwg} which requires an order of magnitude suppression of our $\delta g_A^\ell$. 
Once $\epsilon$ is fixed  by $m_\tau$  to be $10^{-2}$, the correction 
$\delta g_A^\ell$ can be reduced by decreasing
 $\lambda_{1,2}$. To keep $m_\tau$ intact, one has to increase $\lambda_3 \simeq m_\tau /(v \delta g_A)$ to strong-coupling
values of around 10, which renders our calculations unreliable. Another possibility is to produce $m_\tau$ with the direct coupling $\lambda_0$ and generate only $m_\mu$ and $m_e$ through the mixing with heavy fermions, but this would go against the spirit of our approach.
Finally,  $\delta g_A^\ell$ can be reduced by a flavor-universal contribution of additional particles, which do not
affect the flavor structures. We thus conclude that the toy model presented in this section cannot be directly extended to leptons  and some new interactions are necessary to generate $m_\tau$ without inducing too large $\delta g^{\ell}_{L,R}$.

Let us return to the quark sector. After we integrate out at tree level the heavy fermions in the down-left sector, the induced interactions
of order $\epsilon^2$ are
\beq
i {\bar q}_L {\tilde \lambda}^d_1 \left[ \left( {\tilde \lambda}^d_3 {\tilde \lambda}^d_4 \slash{D} +{\rm h.c.}\right) +
{\tilde \lambda}^d_3 \slash{D} {\tilde \lambda}_3^{d \dagger} -\slash{D} {\tilde \lambda}_4^{d \dagger} \frac{\slash{D}}{m_S^2} {\tilde \lambda}^d_4 \slash{D}\right] {\tilde \lambda}_1^{d \dagger} q_L ,
\label{risc}
\eeq
which correspond to the flavor structures shown in \eq{struc4}.  The second interaction contains terms of the form $\slash{D} ( H^\dagger   H  q_{Lj})$, which have the same gauge structure as $\slash{D} q_{Lj}$ and therefore lead to an 
overall rescaling of the kinetic terms $\bar q_{Li} \skew0 \slash{D} q_{Lj} $. On the other hand, the first interaction in \eq{risc} is of the form
\beq
{(\lambda_1^d  \lambda_3^d \lambda_4^{d }  \lambda_1^{d\dagger})_{ij} \over M^4}  
\bar q_{Li} H H^\dagger H i \slash{D} ( H^\dagger  q_{Lj})
      +  {\rm h.c.},
\eeq
involving the different gauge structure $\skew0 \slash{D}( H^\dagger  q_{Lj})$ and thus leading to modified $Z$-fermion couplings.
 In particular, we find  flavor violating $Z\bar d_{L_i}d_{L_j} $ and  $Z\bar u_{L_i}u_{L_j} $  vertices
\begin{equation}
\left( \delta g_L^d\right)_{ij}=
 {\epsilon^2\over 2} (\lambda_1^d  \lambda_3^d \lambda_4^{d }  \lambda_1^{d\dagger} +{\rm h.c.})_{ij} ~,~ 
 \left( \delta g_L^u\right)_{ij}=-{\epsilon^2\over 2} (\lambda_1^u  \lambda_3^u \lambda_4^{u }  \lambda_1^{u\dagger} +{\rm h.c.})_{ij}\;.
\end{equation}
In the right-handed sector, the corresponding flavor violating structures are 
 $\lambda_2^{d\dagger}  \lambda_4^d \lambda_3^{d }  \lambda_2^{d}$ and 
$\lambda_2^{u\dagger}  \lambda_4^u \lambda_3^{u }  \lambda_2^{u}$. 
Such couplings are strongly constrained by flavor physics. The $Z \bar s d$ vertex is constrained
most severely by   $K^+ \rightarrow \pi^+ \nu \bar \nu$ and     $K_L \rightarrow \mu^+ \mu^-$
processes such that the effective coupling $Z \bar s_L d_L$ must be $\lsim 7 \times 10^{-6}$ \cite{Colangelo:1998pm}. The
$Z \bar b s$, $Z \bar b d$  vertices are constrained at the level few$\times 10^{-4}$  by 
  $B_{s,d} \rightarrow \mu^+ \mu^-$ and               $b\rightarrow s l^+ l^-$
 processes \cite{Hiller:2003js}. This implies  that the $Z \bar s d$ vertex should
appear at the $\epsilon^3$ level, while the $Z \bar b s$, $Z \bar b d$ vertices are allowed at the 
$\epsilon^2$ order. This is indeed what we have in our model as $\lambda_1 \lambda_3 \lambda_4
\lambda_1^\dagger$ and 
$\lambda_2^\dagger \lambda_4 \lambda_3 \lambda_2$ 
in both up- and down-sectors have a zero 2$\times2$ block in the upper left corner such
that the (12) transition only appears at order $\epsilon^3$, but the (13) and (23) transitions
already exist at order $\epsilon^2$.
Therefore the flavor physics bounds are satisfied.

We remark that loops with heavy quarks  which produce dipole interactions 
$ {1\over v} \bar d_{Li} \sigma_{\mu \nu} 
d_{Rj} ~ F^{\mu \nu}$
lead to flavor violation in the down sector at order $\epsilon^3$ (or, more precisely, $\epsilon^2$ times a loop factor)
in our model. This is because $\lambda_1^d \lambda_4^{d \dagger} \lambda_2^d=0$ and the lowest
order operators vanish. The resulting contribution to BR$(b \rightarrow s \gamma)$ is small. 
Finally, the induced  quark  EDMs depend strongly on how CP violation is implemented in the model.
In particular, if the  CKM phase is due to  non-removable CP phases  in $\lambda_0$ of \eq{l0}
while the other flavor objects are real in that basis,
the EDM constraints are insignificant.

To conclude, the above model provides an example of how TeV scale new physics 
can generate Higgs-dependent  
structures in the Yukawa couplings. Although this model as it stands is viable only for quarks, while  for leptons
additional flavor-universal interactions are required, 
it shows that it is possible to induce these  flavor  structures without entailing excessive FCNC.

\section{Conclusions}
\label{con}

We have studied the possibility that the SM Yukawa couplings 
are functions of the Higgs field. This 
can explain the hierarchical structure of the fermion masses and
mixing angles in terms of  powers of the ratio between the Higgs VEV and a new
mass scale. We find that this mass scale, characterizing the dynamics of flavor physics, is determined
to be roughly $\sqrt{m_t/m_b}~ v\simeq$~1--2~TeV.

An immediate consequence of this approach is that the Higgs boson
couplings to fermions are drastically modified. The Higgs decay widths
into bottom quarks and into  $\tau$'s are 9 times larger than those in the SM, while 
the one into muons is larger by a factor of  25. The prediction for the Higgs branching 
ratios is shown in fig.~1.
Furthermore, the Higgs couplings violate flavor and this results in observable rates for either
$h\to t\bar c$ or $t \rightarrow h c$, whose branching fractions are expected to be 
of order $10^{-3}$, when kinematically accessible.  Tree-level Higgs exchange also contributes to various FCNC processes, but the effects are consistent with the 
present constraints. The most significant contribution is in the $B$--$\bar B$ system and can be very close to
the current experimental sensitivity. The new flavor dynamics at the TeV scale is within the reach of LHC experiments. 

We have also presented an example of  a possible TeV scale completion of our effective theory.
This model involves heavy vector-like quarks with flavor-universal masses which interact with ordinary quarks
in a flavor conserving way (at leading order). Due to this universality, dangerous FCNC operators
are sufficiently suppressed, while the correct fermion masses and mixings can be generated.

{\bf Acknowledgements.} We are grateful to G.~Degrassi, S.~J\"ager, H.D.~Kim, R.~Rattazzi, M.~Ratz, P.~Slavich and T.~Takeuchi  for discussions. We would also like to thank K.~Babu for pointing out ref.~\cite{Babu:1999me}.


\begin{thebibliography}{99}


\bibitem{Babu:1999me}
  K.~S.~Babu and S.~Nandi,
  Phys.\ Rev.\  D {\bf 62}, 033002 (2000).




\bibitem{Froggatt:1978nt}
  C.~D.~Froggatt and H.~B.~Nielsen,
  Nucl.\ Phys.\  B {\bf 147}, 277 (1979).


\bibitem{Baum:2008qm}
  I.~Baum, G.~Eilam and S.~Bar-Shalom,
  arXiv:0802.2622 [hep-ph].


\bibitem{branco}
  J.~A.~Aguilar-Saavedra and G.~C.~Branco,
  Phys.\ Lett.\  B {\bf 495}, 347 (2000).

\bibitem{lepewwg} 
    [ALEPH Collaboration],
  Phys.\ Rept.\  {\bf 427}, 257 (2006).




\bibitem{Colangelo:1998pm}
  G.~Colangelo and G.~Isidori,
  JHEP {\bf 9809}, 009 (1998).





\bibitem{Hiller:2003js}
  G.~Hiller and F.~Kruger,
  Phys.\ Rev.\  D {\bf 69}, 074020 (2004).



\end{thebibliography}
\end{document}